\documentclass{PoS}

\usepackage{bm}
\usepackage{amsmath}

\title{Traces of non-equilibrium dynamics in relativistic heavy-ion collisions}

\ShortTitle{Traces of non-equilibrium dynamics in relativistic heavy-ion collisions}

\author{\speaker{Pierre Moreau}\\
	Institute for Theoretical Physics, Goethe Universit\"at Frankfurt am Main, Germany\\
	GSI Helmholtzzentrum f\"ur Schwerionenforschung GmbH, Darmstadt, Germany\\
	Frankfurt Institute for Advanced Studies, Goethe Universit\"at Frankfurt am Main, Germany\\
	E-mail: \email{moreau@fias.uni-frankfurt.de}}

\author{Yingru Xu\\
        Department of Physics, Duke University, Durham, NC, USA\\
        E-mail: \email{yx59@phy.duke.edu}}
    
\author{Taesoo Song\\
    	Institut f\"ur Theoretische Physik, Universit\"at Gie{\ss}en, Germany}

\author{Marlene Nahrgang\\
	Department of Physics, Duke University, Durham, NC, USA \\
	SUBATECH UMR 6457 (IMT Atlantique, Universit\'e de Nantes, IN2P3/CNRS), Nantes, France}

\author{Steffen Bass\\
	Department of Physics, Duke University, Durham, NC, USA}

\author{Elena Bratkovskaya \\
	GSI Helmholtzzentrum f\"ur Schwerionenforschung GmbH, Darmstadt, Germany\\
	Institute for Theoretical Physics, Goethe Universit\"at Frankfurt am Main, Germany \\}

\abstract{The impact of non-equilibrium effects on the dynamics of heavy-ion collisions is investigated by comparing a
	non-equilibrium transport approach, the Parton-Hadron-String-Dynamics (PHSD), to a 2D+1 viscous hydrodynamical
	model, which is based on the assumption of local equilibrium and conservation laws. Starting the hydrodynamical model
	from the same non-equilibrium initial condition as in the PHSD, using an equivalent lQCD Equation-of-State (EoS), the
	same transport coefficients, i.e. shear viscosity $\eta$ and the bulk viscosity $\zeta$ in the hydrodynamical model, we
	compare the time evolution of the system in terms of energy density, Fourier transformed energy density, spatial and
	momentum eccentricities and ellipticity in order to quantify the traces of non-equilibrium phenomena. In addition, we
	also investigate the role of initial pre-equilibrium flow on the hydrodynamical evolution and demonstrate its importance
	for final state observables. We find that due to non-equilibrium effects, the event-by-event transport calculations show
	large fluctuations in the collective properties, while ensemble averaged observables are close to the hydrodynamical
	results.}

\FullConference{Critical Point and Onset of Deconfinement\\
	7-11 August, 2017\\
	The Wang Center, Stony Brook University, Stony Brook, NY}

\begin{document}
	
%----------------------------------------------------------------
\section{Introduction}
\label{sec:intro}

In this contribution we summarize the findings from our study in Ref. \cite{Xu:2017pna} where we performed a comparison of two prominent models for the evolution of bulk QCD matter. The first one is a non-equilibrium transport approach, the Parton-Hadron-String-Dynamics (PHSD)~\cite{Cassing:2008sv,PHSD,PHSDrhic}, and the second one a  2D+1 viscous hydrodynamical model, VISHNew~\cite{Song:2007ux,Shen:2014vra} which is based on the assumption of local equilibrium and conservation laws.

Non-equilibrium effects are considered to be strongest during the early phase of the heavy-ion reaction and thus may significantly impact the properties of probes with early production times, such as heavy quarks (charm and bottom hadrons), electromagnetic probes (direct photons and dileptons), and jets. Moreover, some bulk observables, such as correlation functions and higher-order anisotropy coefficients, might also retain traces of non-equilibrium effects	\cite{deSouza:2015ena,DerradideSouza:2011rp,Niemi:2013}. In particular, the impact of the event-by event fluctuations on the collective observables has been studied by Kodama et al.~\cite{DerradideSouza:2011rp}. Based on the comparison of the coarse-grained hydrodynamical evolution with the PHSD dynamics, they find that in spite of large fluctuations on event-by-event basis in the PHSD, the ensemble averages are close to the hydrodynamical limit. A similar behavior  has been pointed out before within the PHSD study in Ref.~\cite{Cassing:2008sv} where a linear correlation of the elliptic flow $v_2$ with the initial spatial eccentricity $\varepsilon_2$ has been obtained for the model study of an expanding partonic fireball (cf. Fig. 7 in Ref.~\cite{Cassing:2008sv}). Such correlations of $v_2$ versus $\varepsilon_2$ are expected in the ideal hydrodynamical case~\cite{Voloshin}. The large event-by-event fluctuations of the charge distributions  has been addressed also in another PHSD study \cite{Konchakovski:2014wqa}.

In the present paper our focus will be on isolating differences in the dynamical evolution of the system that can be attributed to non-equilibrium dynamics. The groundwork laid in this comparative study will hopefully lead to the development of new observables that have an enhanced sensitivity to the non-equilibrium components of the evolution of bulk QCD matter and that will allow us to quantify how far off equilibrium the system actually evolves.

%---------------------------------------------------------------
\section{Description of the models}
\label{sec:models}
\subsection{PHSD transport approach}

The Parton-Hadron-String Dynamics (PHSD) transport approach~\cite{Cassing:2008sv,PHSD,PHSDrhic,Cassing:2008nn} is a microscopic covariant dynamical model for strongly interacting systems formulated on the basis of Kadanoff-Baym equations \cite{Kadanoff1,Kadanoff2} for Green's functions in phase-space representation (in first order gradient expansion beyond the quasi-particle approximation). The approach consistently describes the full evolution of a relativistic heavy-ion collision from the initial hard scatterings and string formation through the dynamical deconfinement phase transition to the strongly-interacting quark-gluon plasma (sQGP) as well as hadronization and the subsequent interactions in the expanding hadronic phase as in the Hadron-String-Dynamics (HSD) transport approach \cite{HSD}. The transport theoretical description of quarks and gluons in the PHSD is based on the Dynamical Quasi-Particle Model (DQPM) for partons that is constructed to reproduce lattice QCD results for the QGP in thermodynamic equilibrium~\cite{Cassing:2008nn,Berrehrah:2016vzw} on the basis of effective propagators for quarks and gluons. The DQPM is thermodynamically consistent and the effective parton propagators incorporate finite masses (scalar mean-fields) for gluons/quarks as well as a finite width that describes the medium dependent reaction rate. For fixed thermodynamic parameters $(T, \mu_q)$ the partonic width's $\Gamma_i(T,\mu_q)$  fix the effective two-body interactions that are presently implemented in the PHSD~\cite{Vitaly}. The PHSD differs from conventional Boltzmann approaches in a couple of essential aspects: 
i) it incorporates dynamical quasi-particles due to the finite width of the spectral functions (imaginary part of the propagators);
ii) it involves scalar mean-fields that substantially drive the collective flow in the partonic phase; 
iii) it is based on a realistic equation of state from lattice QCD and thus describes the speed of sound $c_s(T)$ reliably; 
iv) the hadronization is described by the fusion of off-shell partons to off-shell hadronic states (resonances or strings) and does not violate the second law of thermodynamics;
v) all conservation laws (energy-momentum, flavor currents etc.) are fulfilled in the hadronization contrary to coalescence models; 
vi) the effective partonic cross sections are not given by pQCD but are self-consistently determined within the DQPM and probed by transport coefficients (correlators) in thermodynamic equilibrium. The latter can be calculated within the DQPM or  can be extracted from the PHSD by performing calculations in a finite box with periodic boundary conditions (shear- and bulk viscosity, electric conductivity, magnetic susceptibility etc. \cite{Ozvenchuk:2012kh,Ca13}). 
Both methods show a good agreement.

%-------------------------------------------------------------
\subsection{2D+1 viscous hydrodynamics}

Relativistic hydrodynamical models calculate the space-time evolution of the QGP medium via the conservation equations
\begin{equation}
\partial_\mu T^{\mu\nu} = 0
\label{eq:conservation}
\end{equation}
for the energy-momentum tensor
\begin{equation}
T^{\mu\nu} = e \, u^\mu u^\nu  - \Delta^{\mu\nu} (P + \Pi) + \pi^{\mu\nu},
\label{Tmunu_hydro}
\end{equation}
provided a set of initial conditions for the fluid flow velocity $u^\mu$, energy density $e$, pressure $P$, shear stress tensor $\pi^{\mu\nu}$, and bulk viscous pressure $\Pi$. For our analysis, we use VISH2+1~\cite{Song:2007ux}, which is an extensively tested implementation of boost-invariant viscous hydrodynamics that has been updated to handle fluctuating event-by-event initial conditions~\cite{Shen:2014vra}. We use the method from Ref.~\cite{Liu:2015nwa} for the calculation of the shear stress tensor $\pi^{\mu\nu}$.

This particular implementation of viscous hydrodynamics calculates the time evolution of the viscous corrections through the second-order Israel-Stewart equations \cite{Israel:1979wp, Israel:1976aa} in the 14-momentum approximation, which yields a set of relaxation-type equations~\cite{Denicol:2014vaa}

\begin{equation}
	\tau_\Pi \dot{\Pi} + \Pi =
	-\zeta \theta - \delta_{\Pi\Pi} \Pi\theta + \phi_1 \Pi^2 \nonumber
	+ \lambda_{\Pi\pi} \pi^{\mu\nu} \sigma_{\mu\nu}
	+ \phi_3 \pi^{\mu\nu}\pi_{\mu\nu}
\end{equation}
\begin{equation}
	\tau_\pi \dot{\pi}^{\langle \mu\nu \rangle} + \pi^{\mu\nu} =
	2\eta\sigma^{\mu\nu} + 2\pi_\alpha^{\langle \mu} w^{\nu \rangle \alpha}
	- \delta_{\pi\pi} \pi^{\mu\nu} \theta 
	 + \phi_7 \pi_\alpha^{\langle \mu} \pi^{\nu \rangle \alpha}
	- \tau_{\pi\pi} \pi_\alpha^{\langle \mu}\sigma^{\nu \rangle \alpha} 
	 + \lambda_{\pi\Pi} \Pi \sigma^{\mu\nu} + \phi_6 \Pi \pi^{\mu\nu}.
\end{equation}

Here, $\eta$ and $\zeta$ are the shear and bulk viscosities. For the remaining transport coefficients, we use analytic results derived for a gas of classical particles in the limit of small but finite masses~\cite{Denicol:2014vaa}.

The hydrodynamical equations of motion must be closed by an equation of state (EoS), $P = P(e)$. We use a modern QCD EoS based on continuum extrapolated lattice calculations at zero baryon density published by the HotQCD collaboration~\cite{Bazavov:2014pvz} and blended into a hadron resonance gas EoS in the interval {$110 \le T \le 130$~MeV} using a smooth step interpolation function~\cite{Moreland:2015dvc}. While not identical, this EoS is compatible with the one that the DQPM model (underlying the PHSD approach) is tuned to reproduce.

For our study here, we shall initialize the hydrodynamical calculation with an initial condition extracted from PHSD that provides us with a common starting configuration for both models regarding our comparison of the dynamical evolution of the system.

%\newpage
\section{Non-equilibrium initial conditions}
\label{sec:IC}

In this section we describe the construction of the initial condition for the hydrodynamical evolution from the non-equilibrium PHSD evolution. One should note that PHSD starts its calculation ab initio with two colliding nuclei and makes no equilibrium assumptions regarding the nature of the hot and dense system during the course of its evolution from initial nuclear overlap to final hadronic freeze-out. For the purpose of our comparison we have to select the earliest possible time during the PHSD evolution where the system is in a state in which a hydrodynamical evolution is feasible (e.g. the viscous corrections are already small enough) and generate an initial condition for the hydrodynamical calculation at that time (note that this criterion is less stringent than assuming full momentum isotropization or local thermal equilibrium).

\subsection{Evaluation of the energy-momentum tensor $T^{\mu \nu}$ in PHSD}

The energy-momentum tensor $T^{\mu \nu}(x)$ of an ideal fluid (by removing viscous corrections in Eq. \ref{Tmunu_hydro}) is given by
\begin{equation}
T^{\mu \nu} = (e+P) u^\mu u^\nu -P g^{\mu \nu}
\label{Tmunuideal}
\end{equation}
where $e$ is the energy density, $P$ the thermodynamic pressure expressed in the local rest frame (LRF) and the 4-velocity is $u^\mu =\gamma\ (1,\beta_x,\beta_y,\beta_z)$. Here $\boldsymbol{\beta}$ is the (3-)velocity of the considered fluid element and the associated Lorentz factor is given by $\gamma = 1/\sqrt{1-\boldsymbol{\beta}^2}$.

In order to calculate $T^{\mu \nu}$ in PHSD which fully describes the medium in every space-time coordinate, the space-time is divided into cells of size $\Delta x = 1$~fm, $\Delta y = 1$~fm (which is comparable to the size of a hadron) and $\Delta z \propto 0.5 \times t/\gamma_{NN}$ scaled by $\gamma_{NN}$ to account for the expansion of the system. In each cell, we can obtain $T^{\mu \nu}$ in the computational frame from:
\begin{equation}
T^{\mu \nu}(x) = \sum_{i} \int_0^\infty \frac{d^3p_i}{(2\pi)^3}\ f_i(E_i)\ \frac{p_i^\mu p_i^\nu}{E_i}.
\label{TmunuPHSD}
\end{equation}
where $f_i(E)$ is the distribution function corresponding to the particle $i$, $p_i^\mu $ the 4-momentum and $E_i=p_i^0$ is the energy of the particle $i$. In the case of an ideal fluid, if the matter is at rest ($u^\mu = (1,0,0,0)$) $T^{\mu \nu}(x)$ should only have diagonal components and the energy density in the cell can be identified to the $T^{00}$ component. However, in heavy-ion collisions the matter is viscous, anisotropic and relativistic, thus the different components of the pressure are not equal and it becomes more difficult to extract the relevant information. This especially holds true for the early reaction time at which the initial conditions for hydrodynamical model are taken. In order to obtain the needed quantities ($e,\boldsymbol{\beta}$) from $T^{\mu \nu}$ for the hydrodynamical evolution, we have to express them in the local rest frame (LRF) of each cells of our space-time grid. In the general case, the energy-momentum tensor can always be diagonalized, i.e presented as 

\begin{equation}
T^{\mu \nu}\ (x_\nu)_i = \lambda_i\ (x^\mu)_i = \lambda_i\ g^{\mu \nu}\ (x_\nu)_i,
\label{TmunuL}
\end{equation}

with $i =0,1,2,3$, where its eigenvalues are $\lambda_i $ and the corresponding eigenvectors $(x_\nu)_i$. When $i = 0$, the local energy density $e$ is identified to the eigenvalue of $T^{\mu \nu}$ (Landau matching) and the corresponding time-like eigenvector is defined as the 4-velocity $u_\nu$ (multiplying (\ref{Tmunuideal}) by $u_\nu$):

\begin{equation}
T^{\mu \nu}u_\nu = e u^\mu = (e g^{\mu \nu})u_\nu
\label{Landau_cond}
\end{equation}

using the normalization condition $u^\mu u_\mu =1$. In order to solve this equation, we have to calculate the determinant of the corresponding matrix which is the $4^{\text{th}}$ order characteristic polynomial associated to the eigenvalues $\lambda$:

\begin{equation}
P(\lambda) =
\begin{vmatrix}
T^{00}-\lambda & T^{01} & T^{02} & T^{03} \\
T^{10} & T^{11}+\lambda & T^{12} & T^{13} \\
T^{20} & T^{21} & T^{22}+\lambda & T^{23} \\
T^{30} & T^{31} & T^{32} & T^{33}+\lambda 
\end{vmatrix}
\end{equation}
Having the four solutions for this polynomial, we can identify the energy density being the larger and positive solution, and the 3 other solutions are ($-P_i$) the pressure components expressed in the LRF. To obtain the 4-velocity of the cell, we use (\ref{Landau_cond}) which gives us this set of equations:

\begin{equation}
\left\{ \ \ \ 
\begin{matrix}
(T^{00}-e) + T^{01}X + T^{02}Y + T^{03}Z = 0 \\
T^{10} + (T^{11}+e)X + T^{12}Y + T^{13}Z = 0 \\
T^{20} + T^{21}X + (T^{22}+e)Y + T^{23}Z = 0 \\
T^{30} + T^{31}X + T^{32}Y + (T^{33}+e)Z = 0
\end{matrix}
\right.
\end{equation}
Rearranging these equations, we can obtain the solutions which are actually for the vector $u_\nu = \gamma\ (1,X,Y,Z) = \gamma\ (1,-\beta_x,-\beta_y,-\beta_z)$. To obtain the physical 4-velocity $u^\mu$, we  have to multiply by $g^{\mu \nu} u_\nu = u^\mu$.

\begin{figure}
	\centering
	\includegraphics[width=0.49\textwidth]{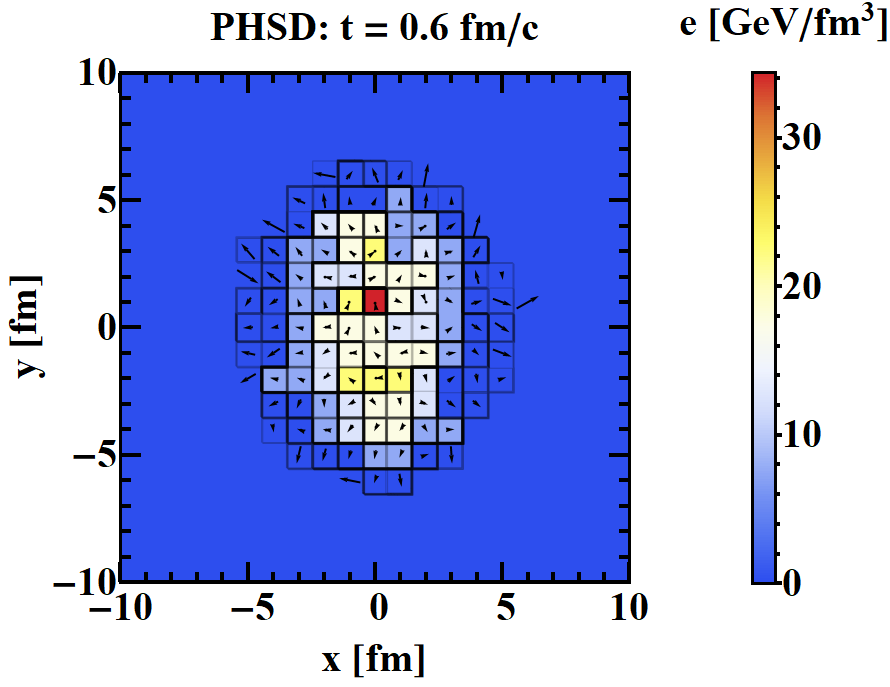}
	\includegraphics[width=0.49\textwidth]{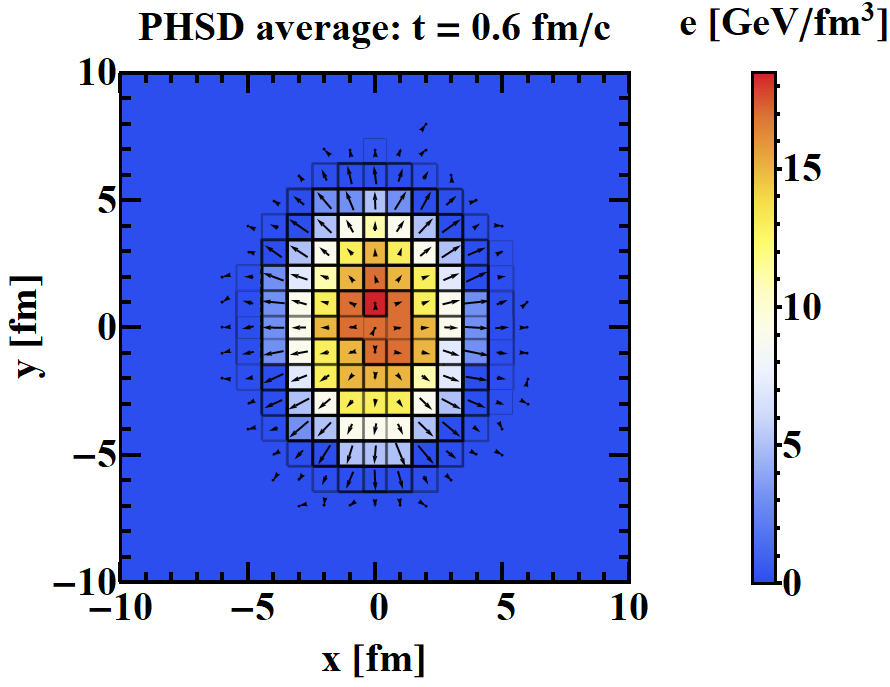}
	\caption{Initial conditions for hydrodynamics: the energy density profiles from a single PHSD event (left panel) and averaged over 100 PHSD events (right panel)
		taken at $t=0.6$ fm/c
		for a peripheral ($b=6$~fm) Au+Au collision at $\sqrt{s_{NN}}=200$ GeV.}
	\label{IC_VISHNU}
\end{figure}

%----------------------------------------------------------
\subsection{PHSD initial conditions for hydrodynamics}

By the Landau matching procedure described above, we can obtain the initial conditions such as the local energy density $e$ and initial flow $\vec{\beta}$ for the hydrodynamical evolution. In the PHSD simulation the parallel ensemble algorithm is used for the test particle method, which has an impact on the fluctuating initial conditions. For a larger number of parallel ensembles (NUM), the energy density profile is smoother since it is calculated on mean-field level by averaging over all ensembles. From a hydrodynamical point of view, gradients should not be too large and some smoothing of the initial conditions is therefore required. Here, we choose NUM$=30$, which provides the same level of smoothing of the initial energy density as in typical PHSD simulations. In fig.~\ref{IC_VISHNU} we show the initial condition at time $t=0.6$ fm/c extracted from a single PHSD event averaged over (NUM=30) parallel events (left panel) and averaged over 100 parallel events (right panel), the color maps represent the local energy density while the arrows show the initial flow at each of the cells. Even though the initial profiles are averaged over NUM$=30$ parallel events, the distribution still captures the feature of event-by-event initial state fluctuations.

\begin{figure}[!]
	\includegraphics[width=1.\textwidth]{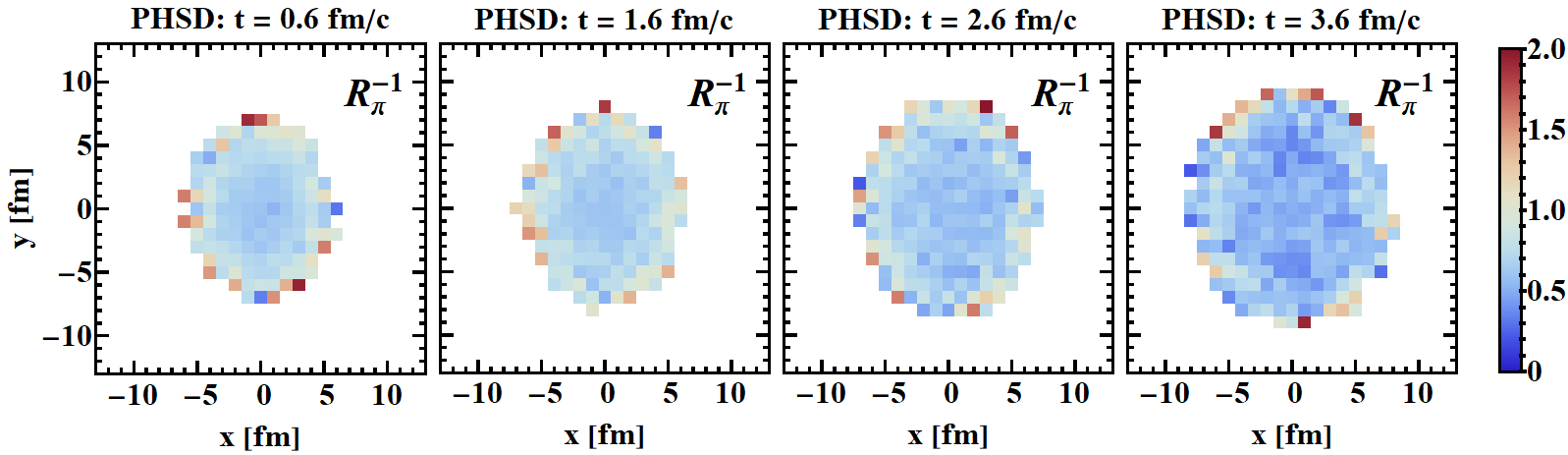}
	\caption{Evolution of the inverse Reynolds number $R^{-1}_\pi$ extracted from PHSD in the transverse plane of a peripheral ($b=6$~fm) Au+Au collision at $\sqrt{s_{NN}}=200$ GeV. Note that $T^{\mu \nu}$ has been averaged over $100$ PHSD events.}
	\label{P_diff}
\end{figure}

In order to justify the choice of initial time $t_0=0.6$~fm, we first took a look in Ref. \cite{Xu:2017pna} at the evolution of the different pressure components in PHSD as a function of time. We have seen that by averaging over many PHSD events, the medium reaches with time a transverse pressure comparable to the isotropic one as given by the lQCD EoS. This statement is of course not valid for a single PHSD event where the pressure components show a much more chaotic behavior and where the high fluctuations in density and velocity profiles indicate that the medium is in a non-equilibrium state, as we will see in the next section. In the pre-equilibrium stage deviations from thermal equilibrium are very large. As a further check, we evaluated the inverse Reynolds number in Fig.~\ref{P_diff} defined as $R^{-1}_\pi = \sqrt{\pi^{\mu \nu}\pi_{\mu \nu}}/P$ which quantifies the applicability of fluid dynamics. One can see that at 0.6 fm/c, the inverse Reynolds number is predominantly below 1 which reinforce this choice for the initialization time of the hydrodynamic simulation.

%---------------------------------------------------------------
\section{Medium evolution: hydrodynamics versus PHSD}

\label{sec:med}

\begin{figure}[!]
	\centering
	{\includegraphics[width=0.49\textwidth]{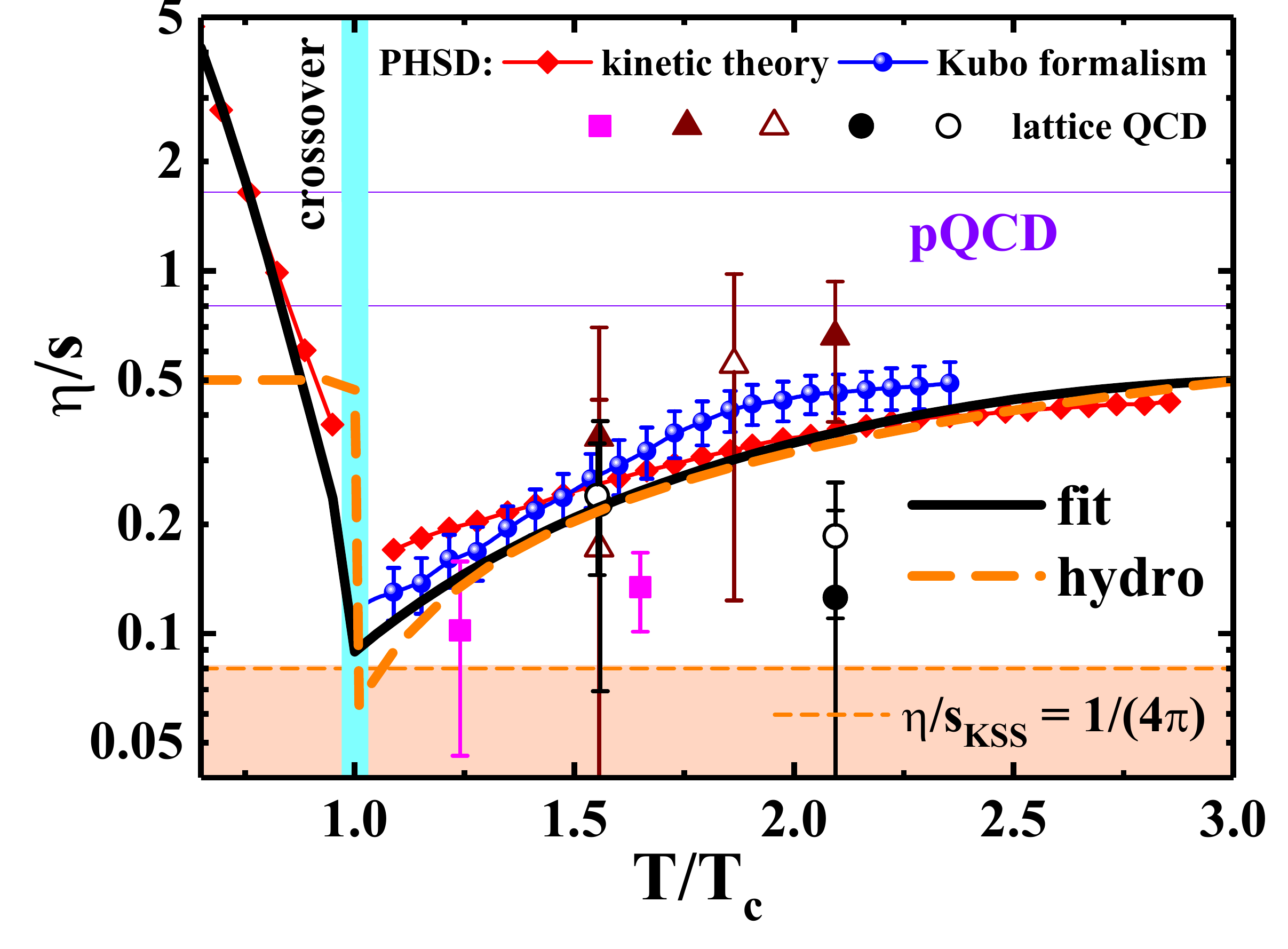}}
	{\includegraphics[width=0.49\textwidth]{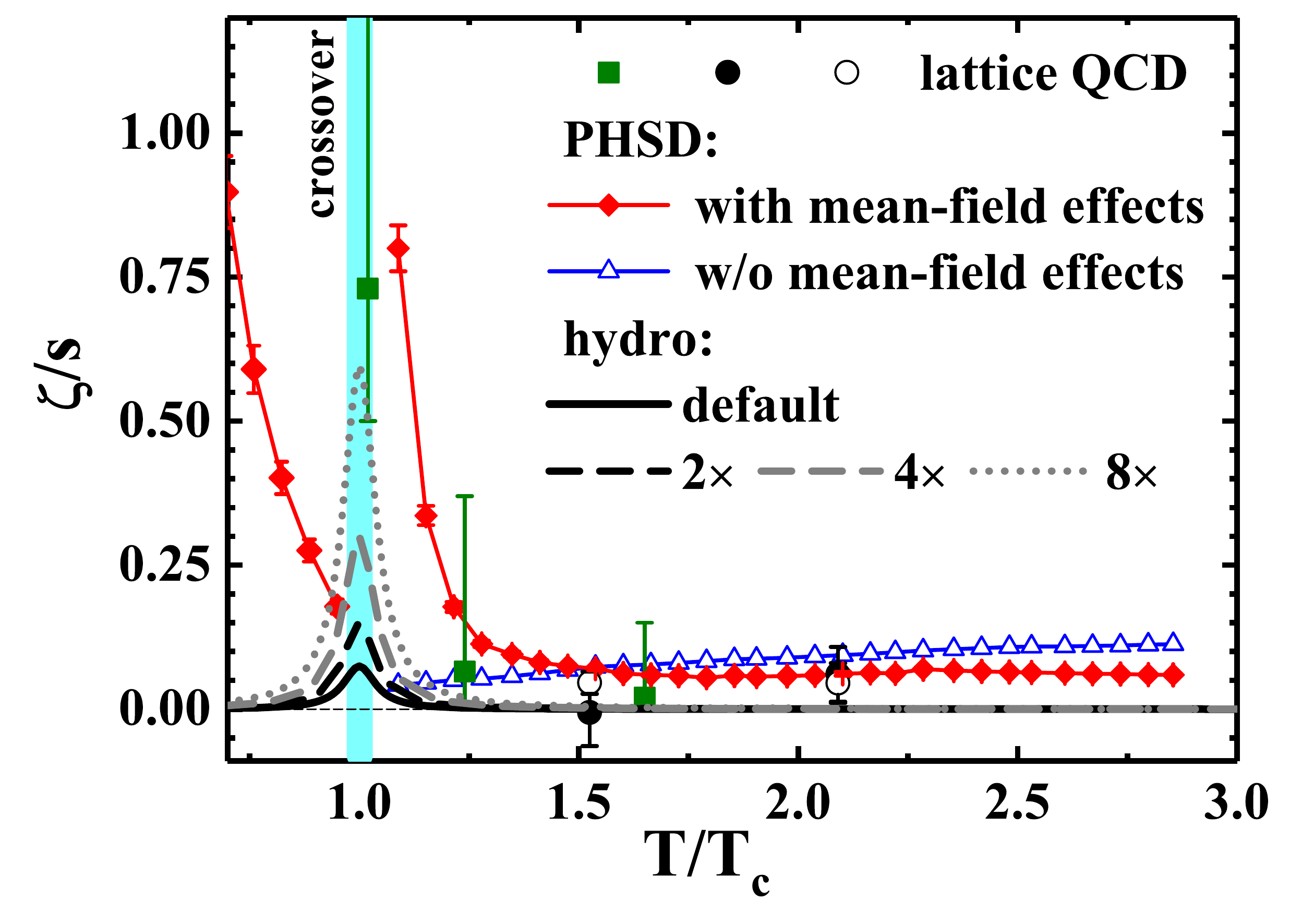}}
	\caption{ (Color online) $\eta/s$ and $\zeta/s$ versus scaled temperature $T/T_C$: 
			\textbf{Left:} 	The symbols indicate the PHSD results of $\eta/s$ from Ref.~\cite{Ozvenchuk:2012kh},
			calculated  using different methods: the relaxation-time approximation (red line$+$diamonds) 
			and the Kubo formalism (blue line$+$dots);
			the black line corresponds to the parametrization of the PHSD results for $\eta/s$. 
			The orange short dashed line demonstrates the Kovtun-Son-Starinets bound \cite{KSS}
			$(\eta/s)_{KSS}=1/(4\pi).$ For comparison, the results from the
			virial expansion approach (green line) \cite{Mattiello} are shown as a function of temperature, too.		
			The orange dashed line is the $\eta/s$ of VISHNU hydrodynamical model that has been recently determined by Bayesian analysis; 
			\textbf{Right:} $\zeta/s$ from PHSD simulation from Ref. \cite{Ozvenchuk:2012kh} and the $\zeta/s$ that is adapted in our hydrodynamical simulations.}
	\label{f1b}
\end{figure}

In this section we compare the response of the hydrodynamical long-wavelength evolution to the PHSD initial conditions with the microscopic PHSD evolution itself. In order to avoid as many biases as possible we apply the temperature-dependent shear viscosity as determined in PHSD simulations~\cite{Ozvenchuk:2012kh} and shown in the left panel of Fig.~\ref{f1b}: the blue and red symbols correspond to $\eta/s$ obtained from the Kubo formalism and from the relaxation time approximation method, respectively. The black line in Fig.~\ref{f1b} shows the parametrization of the PHSD $\eta/s(T)$, which is used in the viscous hydrodynamics for the present study. We note that the parametrized curve is very similar to the recently determined temperature dependence of $\eta/s$ via Bayesian analysis of the available experimental data~\cite{bayesQM2017}. Concerning the bulk viscosity, we decided to use the bulk viscosity that has recently been determined by the Bayesian analysis of experimental data in our hydrodynamical simulations~\cite{bayesQM2017}. In the right panel of Fig.~\ref{f1b} we compare the ratio of bulk viscosity to entropy $\zeta/s$ that is adapted in our hydrodynamical simulations and the one extracted from PHSD simulations. It should be noted that the maximum $\zeta/s$ that hydrodynamical model can handle is much smaller than the bulk viscosity from PHSD simulations, and its effect on the momentum anisotropy will be discussed at the end of this section. 

%------------------------------------------------------------------------------------
\subsection{Space-time evolution of energy density $e$ and velocity $\vec \beta$}

\begin{figure*}
	\centering
	\includegraphics[width=0.99\textwidth]{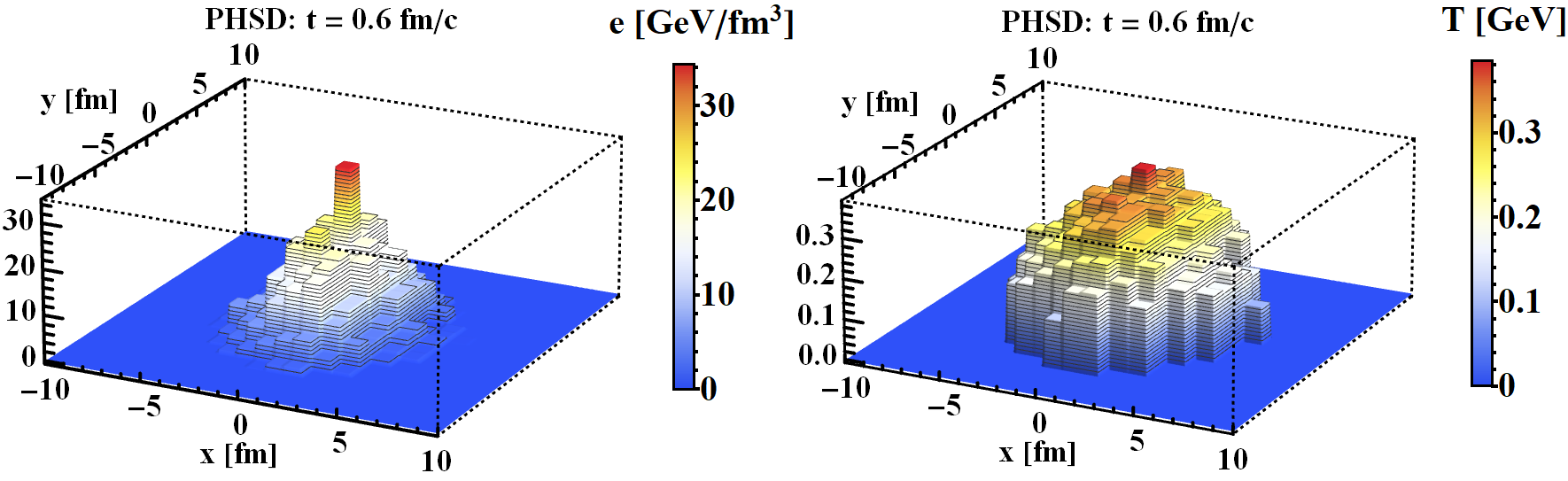}
	\includegraphics[width=0.99\textwidth]{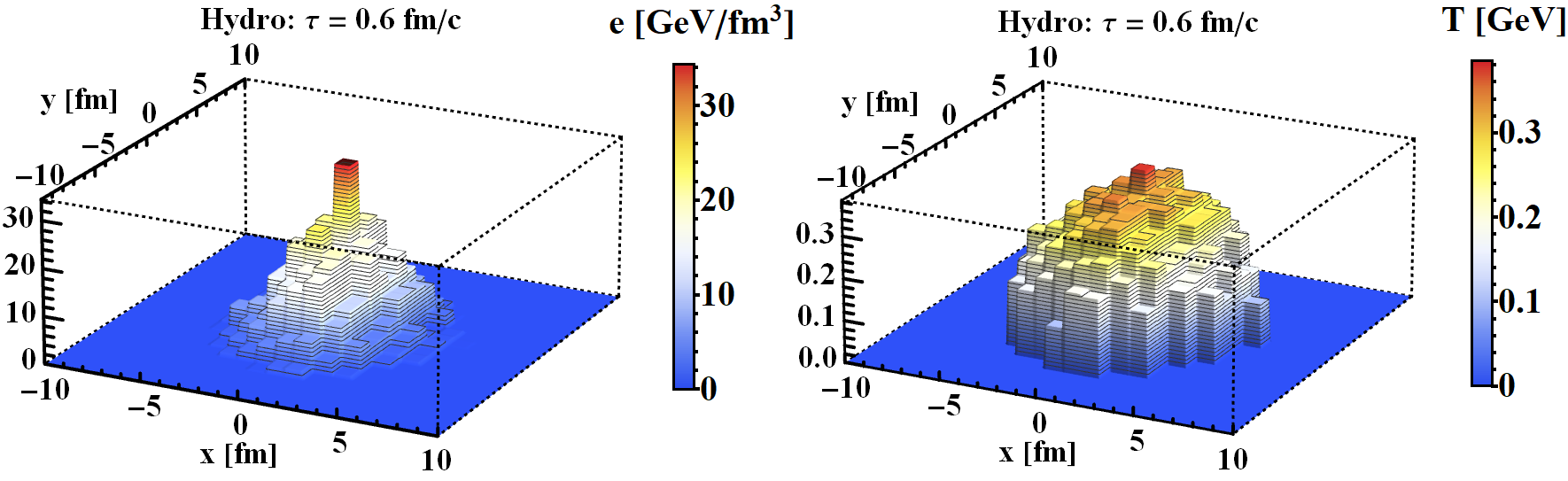}
	\includegraphics[width=0.99\textwidth]{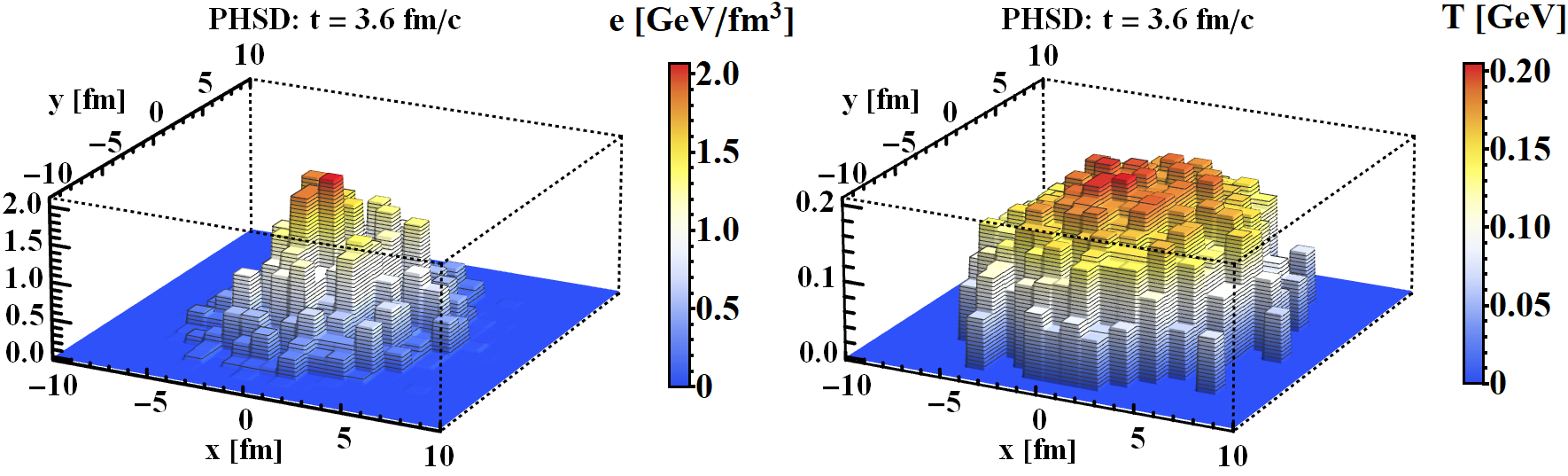}
	\includegraphics[width=0.99\textwidth]{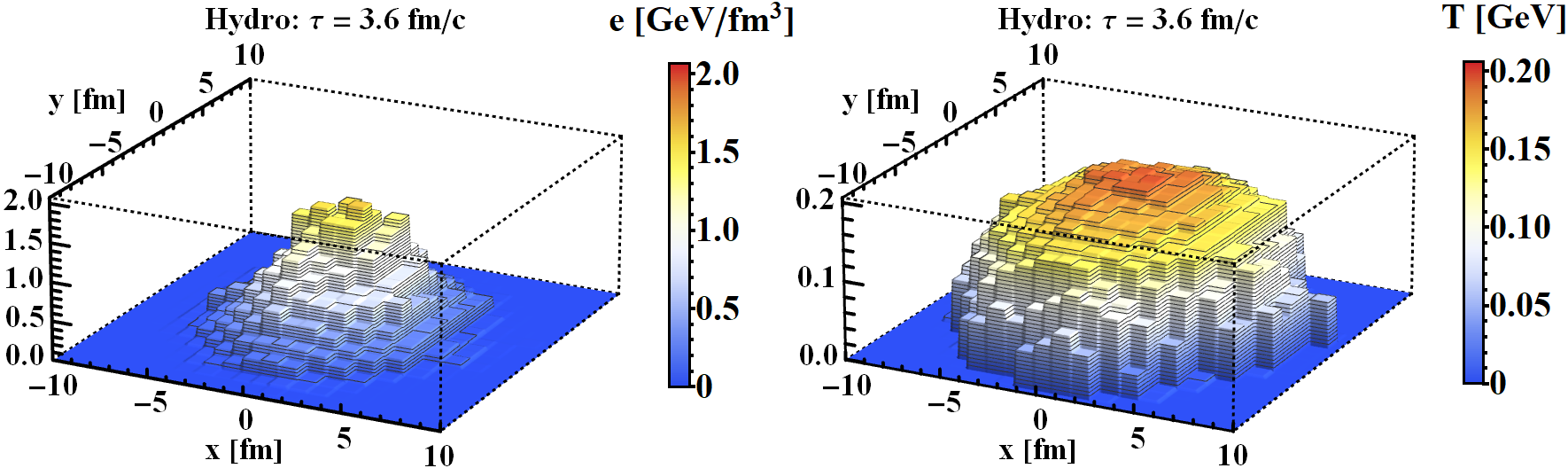}
	\caption{Local energy density $e(x,y,z=0)$ (left) and the corresponding temperature $T(x,y,z=0)$ given by the EoS (right) in the transverse plane from a single PHSD event (NUM=30) and a single hydrodynamical event at different times for a peripheral (b=6fm) Au+Au collision at $\sqrt{s_{NN}}=200$ GeV.}
	\label{Evol_3D_PHSD}
\end{figure*}

Starting with the same initial conditions (as discussed in section~\ref{sec:IC}), the evolution of the QGP medium is now simulated by two different models: the non-equilibrium dynamics model -- PHSD, and hydrodynamics -- (2+1)-dimensional VISHNU.

Fig.~\ref{Evol_3D_PHSD} shows the time evolution of the local energy density $e(x,y,z=0)$ (from $T^{\mu \nu}$) (left) and the corresponding temperature $T$ (right) as calculated using the lQCD EoS in the transverse plane from a single PHSD event (NUM=30) at different times for a peripheral ($b=6$~fm) Au+Au collision at $\sqrt{s_{NN}}=200$ GeV. As seen in figure~\ref{IC_VISHNU} for $t = 0.6$ fm/c, the energy density profile is far from being smooth. Note also that the energy density decreases rapidly as the medium expands in the transverse and longitudinal directions. By converting the energy density to the temperature given by the lQCD EoS, we can see that the variations are less pronounced in that case. In particular for the energy density at later times one can already observe a significant smoothing compared to the PHSD evolution. A comparison of the two medium evolutions shows distinct differences: in PHSD the energy density retains many small hot spots during its evolution due to its spatial non-uniformly. In hydrodynamics, the initial hot spots of energy density quickly dissolve and  the medium becomes much smoother with  increasing time. We attribute these differences directly to the non-equilibrium nature of the PHSD evolution. Moreover, as a result of the initial spatial anisotropy, the pressure gradient in $x$-direction is larger than that in $y$-direction, resulting in a slightly faster expansion in $x$-direction.

%------------------------------------
\subsection{Time evolution of the spatial and momentum anisotropy}

Much interest is given to the medium's response to initial spatial anisotropies. For the hydrodynamical models the spatial anisotropies lead to substantial collective flow, measured by Fourier coefficients of the azimuthal particle spectra.  Initial spatial gradients are transformed into momentum anisotropies via hydrodynamical pressure. While experimentally only the final state particle spectra are known, models for the space-time evolution of the medium can give insight into the evolution of the spatial and the momentum anisotropy. For hydrodynamical models the latter is directly related to the elliptic flow $v_2$. Similar statements apply to the transport models where the initial spatial anisotropies are converted to momentum anisotropies~\cite{Cassing:2008sv}.

The spatial anisotropy of the matter distribution is quantified by the eccentricity coefficients $\epsilon_n$ defined as
\begin{equation}
\epsilon_n \exp({i n \Phi_n}) = - \frac{\int rdr d\phi r^n \exp({in\phi}) e(r, \phi)}{\int rdr d\phi r^n e(r, \phi)}
\end{equation}
where $e(r, \phi)$ is the local energy density in the transverse plane. 

The second-order coefficient $\epsilon_2$ is also called ellipticity and to leading order the origin of the elliptic flow $v_2$. It can be simplified to
\begin{equation}
\epsilon_2 = \frac{\sqrt{\{r^2 \cos(2\phi)\}^2 + \{r^2 \sin(2\phi)\}^2}}{\{r^2\}}
\end{equation}
where $\{...\} = \int dxdy (...) e(x,y)$ describes an event-averaged quantity weighted by the local energy density $e(x,y)$~\cite{Qiu:2011iv}.

\begin{figure}
	\centering
	\includegraphics[width=0.48\textwidth]{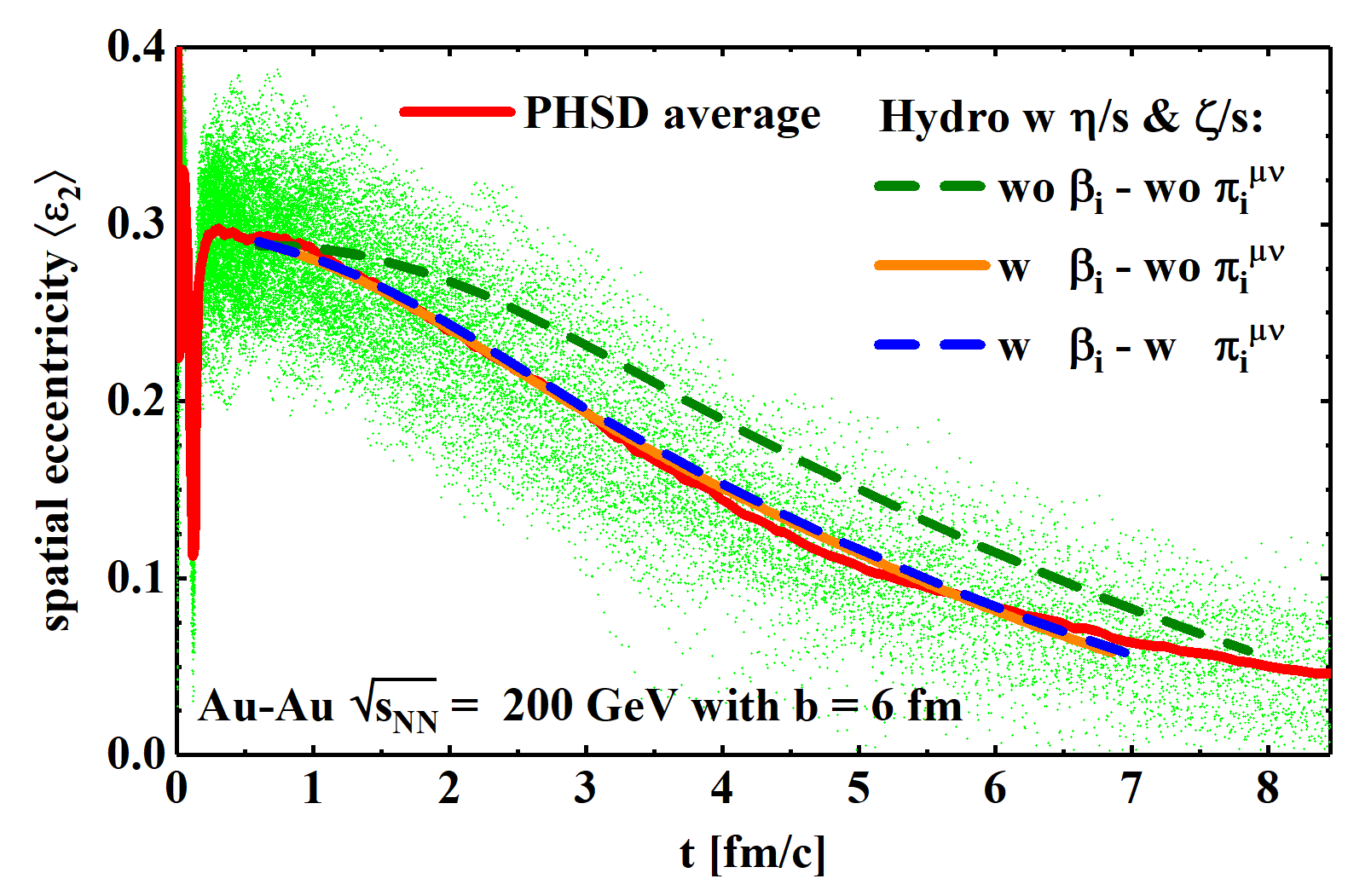}
	\caption{(Color online) Event-by-event averaged spatial eccentricity $\epsilon_2$ of 100 PHSD events and 100 VISHNU events with respect to proper time, for a peripheral Au+Au collision ($b=6$~fm) at $\sqrt{s_{NN}}=200$ GeV. The green dots show the distribution of each of the 100 PHSD events used in this analysis. The solid red line is the average over all the green dots. The blue, yellow and black line correspond to hydrodynamical evolution taking different initial condition scenarios.}
	\label{fig:Eccenx}
\end{figure}

\begin{figure*}
	\centering
	\includegraphics[width=1.0\textwidth]{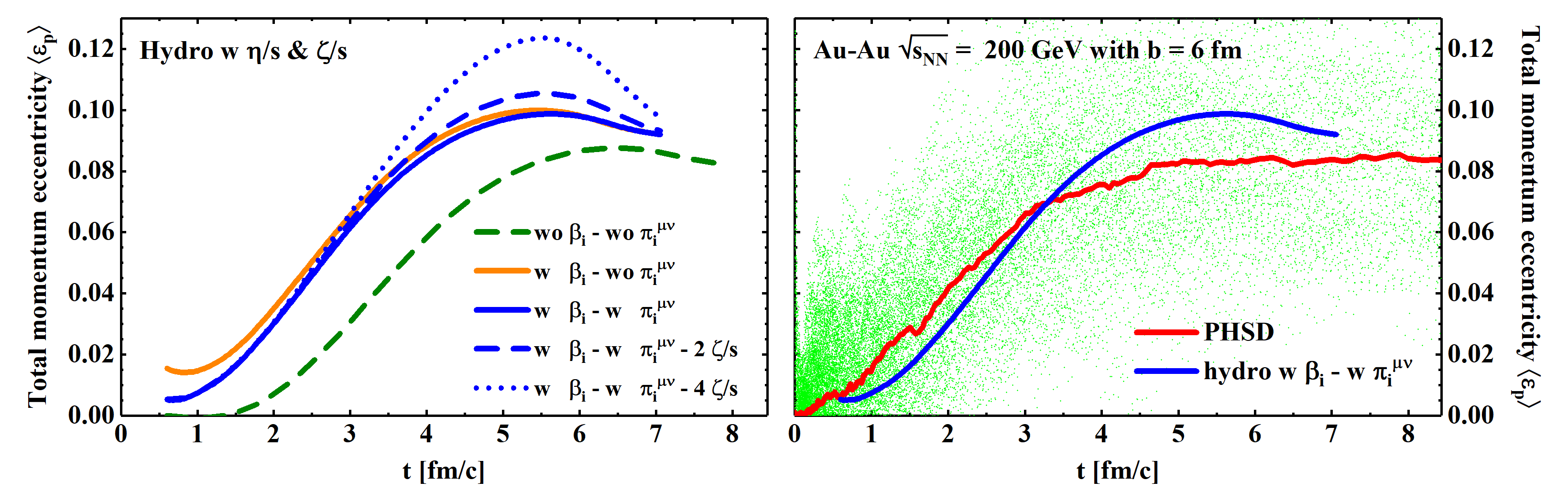}
	\caption{(Color online) Event-by-event averaged total momentum anisotropy of 100 PHSD events and 100 VISHNU events with respect to proper time, for a peripheral Au+Au collision ($b=6$ fm) at $\sqrt{s_{NN}}=200$ GeV. \textbf{Left:} the total momentum eccentricity of hydrodynamical evolution for  different initial scenarios, as well as different bulk viscosity adapted in the hydrodynamical simulation. \textbf{Right:} comparison of the total momentum eccentricity from PHSD events compared with the standard hydrodynamical events. The green dots show the distribution of each of the 100 PHSD events used in this analysis. The solid red line is an average over the green dots. The black line corresponds to the standard hydrodynamical evolution taking the 100 initial conditions which are generated from PHSD events.}
	\label{fig:Eccenp}
\end{figure*}

The importance of event-by-event fluctuations in the initial state has been realized in particular for higher-order flow harmonics but also as a contribution to the elliptic flow and has been extensively investigated both experimentally and theoretically~\cite{Miller:2003kd,Alver:2006wh,Alver:2010gr}. As shown earlier, the PHSD model naturally produces initial state fluctuations due to its microscopic dynamics. We therefore apply event-by-event hydrodynamics and all subsequent quantities are averaged over many events.

In Fig.~\ref{fig:Eccenx} we show the time evolution of the ellipticity $\langle \epsilon_2\rangle$ for both medium descriptions. For the PHSD simulations we observe large oscillations in $\langle \epsilon_2\rangle$ at the beginning of the evolution due to the initialization geometries and formation times. After sufficient overlap of the colliding nuclei at the initial time $\tau_0$ the average  $\langle \epsilon_2\rangle$  is stabilized in PHSD. There are, however, still significant event-by-event fluctuations of this quantity at later times and strong variations between individual events.

In contrast, in a single hydrodynamical event $\epsilon_p$ deviates from the average, but remains a smooth function of time. Due to the faster expansion in $x$-direction the initial spatial anisotropy decreases during the evolution for both medium descriptions. However, the spatial anisotropy decreases faster when initial pre-equilibrium flow  $\beta_i$ (extracted from the early PHSD evolution) is included in the hydrodynamical evolution. In this case, the time evolution of the event-by-event averaged spatial anisotropy is very similar in PHSD and in hydrodynamics. Initializing with the shear-stress tensor $\pi_i^{\mu\nu}$ may have slight effects on the spatial eccentricity but not large enough to be visible.

A similar feature is also seen in the evolution of the momentum ellipticity, which is directly related to the integrated elliptic flow $v_2$ of light hadrons. The total momentum ellipticity is determined from the energy-momentum tensor as~\cite{Kolb:2003dz, Liu:2015nwa}:
\begin{equation}
\epsilon_p = \frac{\int dx dy (T^{xx} - T^{yy})}{\int dx dy (T^{xx} + T^{yy})}
\label{epsprime}
\end{equation}
Here the energy-momentum tensor includes the viscous corrections from  $\pi^{\mu\nu}$ and $\Pi$.

In the left panel of Fig.~\ref{fig:Eccenp} we show the time evolution of the event-by-event averaged $\langle\epsilon(p)\rangle$ for the hydrodynamical medium description with and without pre-equilibrium flow in the initial conditions. Including the initial flow leads to a finite momentum anisotropy  at $\tau_0$ which subsequently increases as the pressure transforms the spatial anisotropy in collective flow. Consequently, $\epsilon_p$ is larger than in the scenario without initial flow throughout the entire evolution of the medium and an enhanced elliptic flow can be expected. We also see that for an enhanced bulk viscosity around $T_c$ the momentum anisotropy develops a bump at later times, which is more pronounced for larger bulk viscosity.

In the right panel of Fig.~\ref{fig:Eccenp} the hydrodynamical simulation is compared to the results from PHSD, again for event-by-event averaged quantities and the event-by-event fluctuations indicated by the spread of the cloud. The PHSD momentum eccentricity is constructed by Eq.~(\ref{epsprime}) where $T^{\mu \nu}$ is evaluated from Eq.~(\ref{TmunuPHSD}). It can be observed that before $\tau_0$ the averaged momentum anisotropy in PHSD develops continuously during the initial stage, before it reaches the value which is provided in the initial conditions for hydrodynamics. Despite the seemingly large bulk viscosity, as discussed in the beginning of this section, the momentum anisotropy in PHSD does not show any hint of a bump like in the hydrodynamical calculation. The response to intrinsic bulk viscosity in a microscopic transport model does not seem to be as strong as in hydrodynamics.

\section{Summary} 
\label{sec:sum}
In this work, we have compared two commonly used descriptions of the evolution of a QGP medium in heavy-ion collisions, the microscopic off-shell transport approach PHSD and a macroscopic hydrodynamical evolution. Both approaches give an excellent agreement with numerous experimental data, despite the very different assumptions inherent in these models. In PHSD, quasi-particles are treated in off-shell transport with thermal masses and widths which reproduce the lattice QCD equation of state and are determined from parallel event runs in the simulations. Hydrodynamics assumes local equilibrium to be reached in the initial stages of heavy-ion collisions and transports energy-momentum and charge densities according to the lattice QCD equation of state and transport coefficients such as the shear and bulk viscosity. We have tried to match the hydrodynamical evolution as closely as possible to these quantities as obtained within PHSD: 
\begin{enumerate}
	\item by construction the equation of state in PHSD is compatible with the lQCD equation of state used in the hydrodynamical evolution
	\item a new Landau-matching procedure was used to determine initial conditions for hydrodynamics from the PHSD simulation,
	\item the hydrodynamical simulations utilize the same $\eta/s(T)$ as obtained 
	within PHSD and
	\item different bulk viscosity parameterizations have been introduced in the hydrodynamical simulation that resemble to those obtained in (dynamical) quasi-particle models, which are the basis for PHSD simulations.
\end{enumerate}

In general we find that the ensemble averages over  PHSD events follow closely  the hydrodynamical evolution. The major differences between the macroscopic near-(local)-equilibrium and the microscopic off-equilibrium dynamics can be summarized as:
\begin{enumerate}
	\item A strong short-wavelength spatial irregularity in PHSD at all times during the evolution versus a fast smoothing of initial irregularities in the hydrodynamical evolution such that only global long-wavelength structures survive. These structures have been calculated on the level of the fluid velocity and energy density and quantified in terms of the Fourier modes of the energy density in Ref. \cite{Xu:2017pna}. Due to the QCD equation of state the irregularities imprinted in the temperature are smaller than in the energy density itself.
	\item The hydrodynamical response to changing transport coefficients, especially the bulk viscosity, has a strong impact on the time evolution of the momentum anisotropy. In PHSD these transport coefficients can be determined but remain intrinsically linked to the interaction cross sections. Although there are indications for a substantial bulk viscosity in PHSD, it does not show the same sensitivity to the momentum space anisotropy  as in hydrodynamical simulations.
	\item Event-by-event fluctuations might be of similar magnitude in quantities like the spatial and momentum anisotropy but while they remain smooth functions of  time in hydrodynamics significant variations are observed within in a single event in PHSD as a function of time.
\end{enumerate}

After having gained an improved understanding of the similarities and differences in the evolution of bulk QCD matter between the non-equilibrium PHSD and the equilibrium hydrodynamic approach, we plan to utilize our insights in future projects regarding the development of observables sensitive to non-equilibrium effects and the impact these effects may have  on hard probe observables.

\section*{Acknowledgements}

We appreciate fruitful discussions with J. Aichelin, W. Cassing, P.-B. Gossiaux, T. Kodama. This work in part was supported by the LOEWE center HIC for FAIR as well as by BMBF and DAAD. The computational resources have been provided by the LOEWE-CSC. PM acknowledges support by the Deutsche Forschungsgemeinschaft (DFG) through the grant CRC-TR 211 ``Strong-interaction matter under extreme conditions. SAB, MG and YX acknowledge support by the U.S.\ Department of Energy under grant no.~DE-FG02-05ER41367.

%--------------------------------------------------------------------------------------

\end{document}